\documentclass[10pt,letterpaper]{article}

\usepackage{graphicx}
\usepackage{hyperref}
\usepackage{amsmath}
\usepackage{algorithm}
\usepackage{algorithmic}
\usepackage{enumitem}

\usepackage{authblk}

\usepackage{cogsci}
\usepackage{pslatex}
\usepackage{apacite}

\title{Complexity Reduction in the Negotiation of New Lexical Conventions}

\author[1, 2]{\large \bf William Schueller (william.schueller@inria.fr)}
\author[3, 4, 5]{\large \bf Vittorio Loreto}
\author[2]{\large \bf Pierre-Yves Oudeyer}

\affil[1]{University of Bordeaux, Bordeaux, France}
\affil[2]{INRIA Bordeaux Sud-Ouest/ Ensta ParisTech : Flowers Project-team, Bordeaux, France}
\affil[3]{SONY Computer Science Lab, Paris, France}
\affil[4]{Physics Dpt., Sapienza University of Rome, Rome, Italy}
\affil[5]{Complexity Science Hub Vienna (CSHV), Vienna, Austria}

\begin{document}

\maketitle

\begin{abstract}

In the process of collectively inventing new words for new concepts in a population, conflicts can quickly become numerous, in the form of synonymy and homonymy. Remembering all of them could cost too much memory, and remembering too few may slow down the overall process. Is there an efficient behavior that could help balance the two? The Naming Game is a multi-agent computational model for the emergence of language, focusing on the negotiation of new lexical conventions, where a common lexicon self-organizes but going through a phase of high complexity. Previous work has been done on the control of complexity growth in this particular model, by allowing agents to actively choose what they talk about. However, those strategies were relying on ad hoc heuristics highly dependent on fine-tuning of parameters. We define here a new principled measure and a new strategy, based on the beliefs of each agent on the global state of the population. The measure does not rely on heavy computation, and is cognitively plausible. The new strategy yields an efficient control of complexity growth, along with a faster agreement process. Also, we show that short-term memory is enough to build relevant beliefs about the global lexicon.

\end{abstract}

\section{Motivations}
Lexical conventions constitute an important element of social interactions. They can emerge, evolve, or be learnt within a population, without necessarily having a centralized control. In other words, they can be negotiated through local interactions between individuals. In practice, this happens continuously in human societies, being the spread of new words and conventions, the acquisition of those conventions by infants or other learners, or even the emergence of new forms of communication. Despite the high complexity of the processes involved, humans deal with these issues quite efficiently.

Learning of high complexity tasks in individuals can in general be facilitated by an active control of the complexity of learning situations , often driven by intrinsic motivation, like for example maximization of the learning progress \cite{Gottlieb2013,baldassarre2013,barto2013,oudeyer07}. This type of mechanism is also argued to be an evolutionary advantage for cognitive abilities \cite{oudeyer_evolution}, and can also be found in lexicon acquisition at the individual level \cite{Partridge2015}. But does it have a significant impact on population-wide learning and conventions negotiation dynamics?

The Naming Game \cite{steelskaplan1998_1,Wellens2012,Loreto2011,ke2002self} is an adapted framework to test this hypothesis. It is a class of multi-agent models of language emergence and evolution, where pairs of randomly selected individuals try to communicate by referring to some pre-defined meanings using words. At the beginning, they do not share any convention about word-meaning associations. Through repeated decentralized interactions, a common lexicon self-organizes. However, the process can be slow and pass through a high-complexity phase where agents memorize a lot of conflictual information, in the form of synonyms and homonyms.

It has already been shown that active learning mechanisms can increase convergence speed towards a shared lexicon in different language emergence models
\cite{Cornudella2015,Schueller2016}. The main idea behind those mechanisms is to allow agents to actively choose the topic of their communication, based on information collected during their past interactions and driven by control of complexity growth. However, the algorithms used so far are based on ad hoc heuristics, constrained interaction scenarios and can depend heavily on fine-tuning of parameters themselves depending on population size and number of words and meanings.

In previous work, an approximation of the global state is built by each agent using the information of past interactions, in the form of an average vocabulary of the population \cite{oliphant1997,devylder2007}. Is it possible to design a new principled algorithm for an active topic choice based on such a representation? Could decisions be driven by both the compatibility of an agent's own lexicon with this average vocabulary, and a reduction of both their complexities? Such an algorithm should rely on a time scale for the memory of past interactions: Indeed, in the case of uncentralized negotiation of a lexicon, conflictual conventions will necessarily appear and have to be forgotten in order to converge to a functional global vocabulary. Remembering them could slow down the self-organization process.

In this work, we define a principled measure of correlation between an agent's lexicon and a local approximation of the average lexicon of the population. We build a strategy driven by the maximization of this value without being computationally hard, to be cognitively plausible. We study and discuss the impact of this strategy on convergence time and complexity growth, depending on a time scale used for memory.
\section{Methods}
\subsection{The Naming Game\label{NG}}

\begin{figure}[h!]
\begin{center}
\includegraphics[width=7cm]{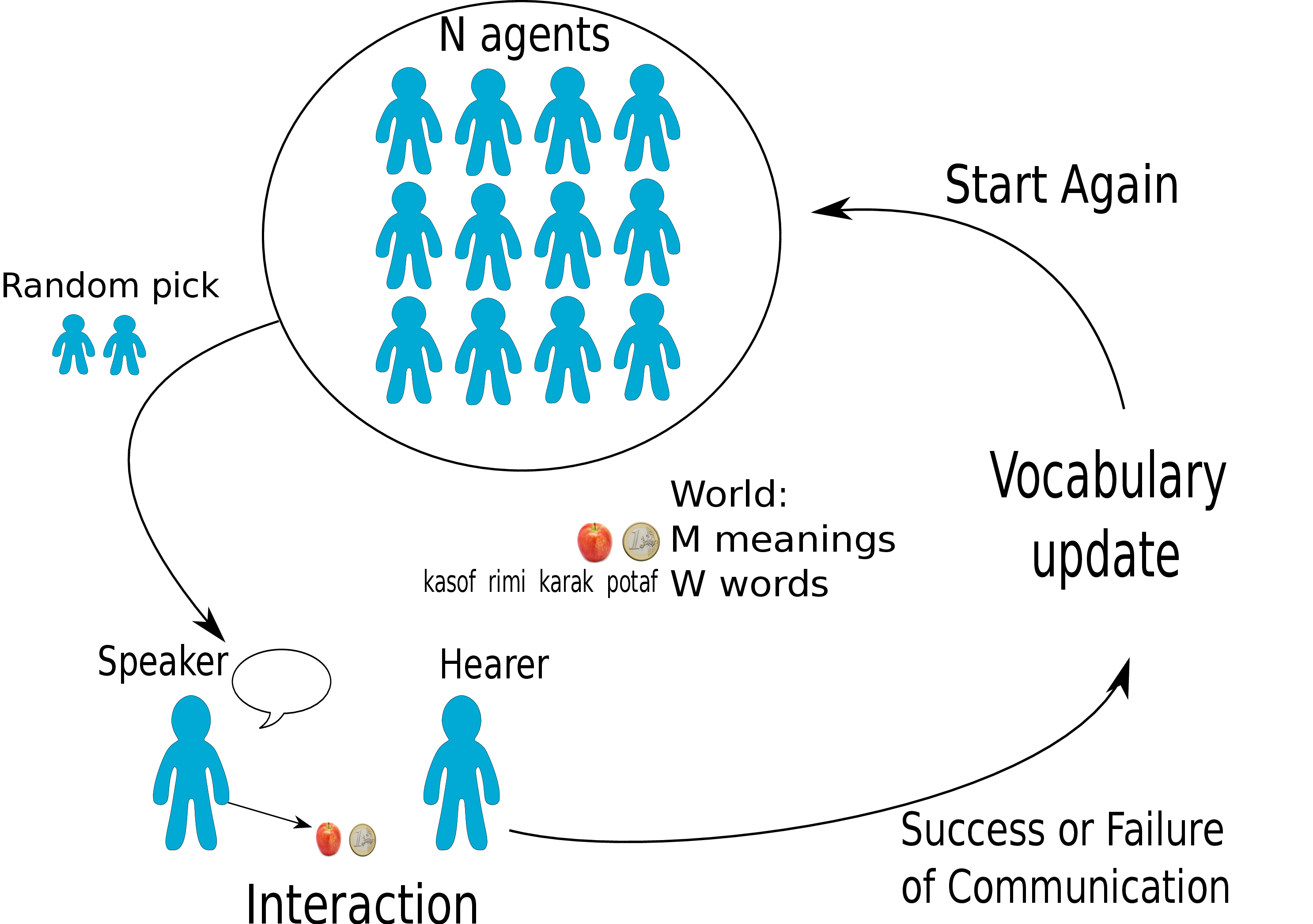}
\end{center}
\caption{{\footnotesize Illustration of the Naming Game model. Out of a population of simulated agents, two are picked and try to communicate, using/inventing words to refer to meanings. After repeated such interactions, a common lexicon self-organizes. In this example, there are $M$=$2$ possible meanings, $W$=$4$ possible words and $N$=$12$ agents.}}
\label{fig_NG}
\end{figure}

We define here precisely the Naming Game model that we used (see fig.\ref{fig_NG} for an overview). We need to explicit:
\begin{itemize}[noitemsep,nolistsep]
\item The interaction scenario itself
\item How agents represent their lexicon
\item How they update their lexicon at the end of each interaction
\end{itemize}
It is a simple modification of the standard Naming Game scenario \cite{Loreto2011,Wellens2012}.

\subsubsection{Interaction process \label{interactions}}

We re-use a previously defined interaction process called \textit{Speaker's Choice} \cite{Schueller2016}. It allows one of the interacting agents, called the speaker, choose actively the topic of the interaction.

\begin{figure}[h!]
\begin{center}
\includegraphics[width=7cm]{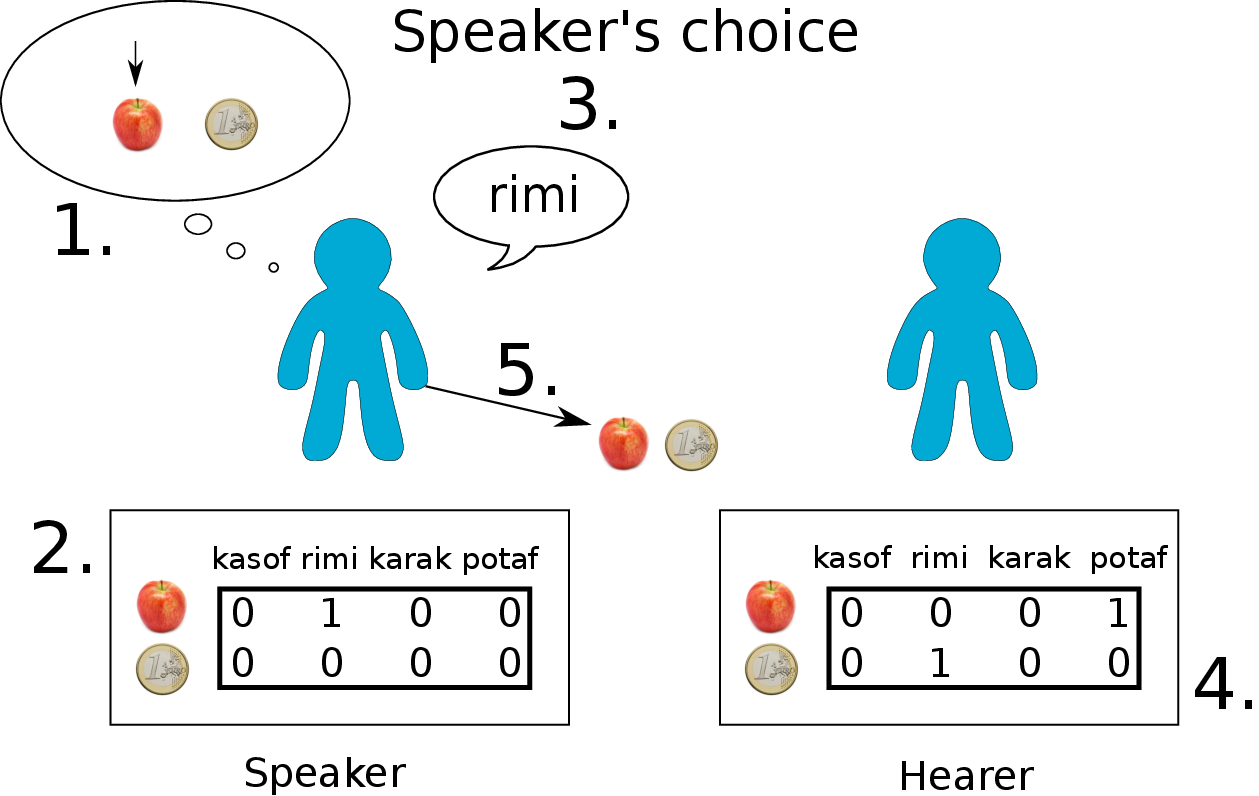}
\end{center}
\caption{{\footnotesize Interaction process: Beforehand, 2 individuals have been randomly selected among a population, an designated as speaker (S) and hearer (H). 1. S chooses a topic, 2. S checks its vocabulary to find or invent an associated word, 3. S utters the word, 4. H guesses the intended meaning, 5. S indicates the intended meaning.}}
\label{fig_speakerschoice}
\end{figure}

Each interaction involves two agents, that are picked randomly from the population. One of them is assigned the role \textit{speaker}, and the other the role \textit{hearer}. The speaker chooses a topic and picks up a word for this topic. If it does not have a word associated so far to the meaning used as topic, it just invents a new meaning-word association. It utters this word, which is interpreted by the hearer as a meaning, if it knows this word. If the interpreted meaning is the same as the topic, i.e. the meaning intended by the speaker, the communication is considered successful. Otherwise, it is considered a failure. See fig.\ref{fig_speakerschoice} for a detailed illustration of the interaction process.
\subsubsection{Vocabulary Representation \label{voc_rep}}

Vocabularies, or lexicons, are a set of associations between meanings and words. In this work, we consider only a finite set of meanings $M$ and a finite set of words $W$. In this context, vocabularies can be represented as associations matrices, where each row corresponds to a meaning, and each column to a word. This representation has been extensively used in related work \cite{oliphant1997,steelskaplan1998_1,ke2002self}. Two parts of the lexicon are distinguished, the coding or production part, which maps a meaning to a set of words weighted by probabilities of usage, and a decoding or interpretation part, mapping a word to a set of meanings that can be interpretated from this word, also weighted by probabilities.

We represent the vocabulary of an agent $A$ as a matrix $V(A)$ of size $M\times W$, with values of $1$ for each word-meaning association used by the agent. Each agent starts with an empty vocabulary, a matrix filled with zeros. The coding matrix $V^c(A)$ and decoding matrix $V^d(A)$ are derived from $V(A)$ by normalizing respectively over rows and columns:
\begin{equation}
V^c(A)_{mw} = \frac{V(A)_{mw}}{\sum\limits_{w'}V(A)_{mw'}} \hspace{0.8cm} V^d(A)_{mw} = \frac{V(A)_{mw}}{\sum\limits_{m'}V(A)_{m'w}}
\end{equation}
Normalization factors are used only if $V(A)_{mw} \neq 0$.
In practice, when coding a meaning $m$, a word $w_i$ is sampled using the distribution $\left( V^c(A)_{mw}\right)_{w \in W} $. When decoding a word $w$, a meaning $m_j$ is interpreted, sampled from the distribution $\left( V^d(A)_{mw}\right)_{m \in M}$. In our case, these distributions are uniform either on the set of words associated to $m$ for coding, or on the set of meanings associated to $w$ for decoding. Those 2 sets change over time, during the vocabulary update.

\subsubsection{Vocabulary Update Policy \label{vocupdate}}

At the end of each interaction, each agent takes into account the result of the interaction by modifying its lexicon. There exists various policies that have been described and studied in previous work \cite{Wellens2012}. We are using the one called Minimal Naming Game.

\begin{figure}[h!]
\begin{center}
\includegraphics[width=7cm]{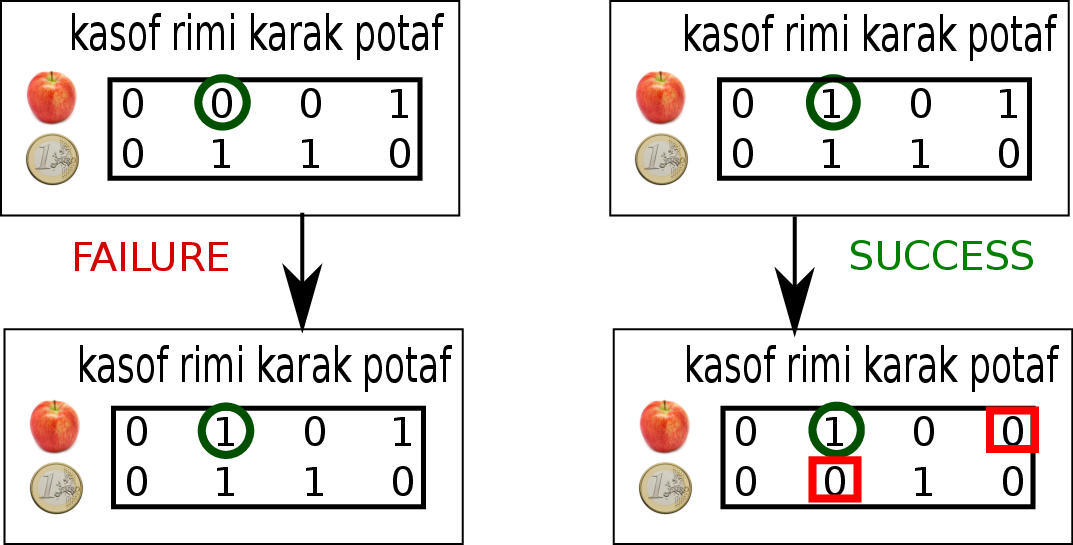}
\end{center}
\caption{{\footnotesize Vocabulary update (Minimal NG). Failure (when the hearer interpreted the word as another meaning than the topic): the word-meaning association used by the speaker is added to the hearer's vocabulary. The speaker adds it as well if it was just invented. Success (when the hearer interpreted correctly the word as the topic): both hearer and speaker remove synonyms/homonyms in conflict with the word-meaning association used during the interaction. In both examples the topic is the apple, and the word \textit{rimi}.}}
\label{fig_success}
\end{figure}

In this policy, updates work this way: when the communication fails, both agents add the used word-meaning association (meaning used as a topic by the speaker, and word uttered by the speaker) to their lexicon, and do nothing if they already had it. If the communication is successful, not only do they add this association to their respective lexicons, they also remove any conflicting synonyms and homonyms. See fig.\ref{fig_success} for an illustration of the update policy in both cases.

Typically, among existing policies, Minimal NG and another one called Basic Lateral Inhibition are used: they are more realistic as they allow synonymy/homonymy and yield faster agreement. Moreover, Minimal NG has been shown to yield similar dynamics as Basic Lateral Inhibition, yet being simple and not depending on any parameter, while the latter depends on 3. This is the reason why we are using the Minimal NG as vocabulary update policy.

\subsection{Measures}

The self-organization process happening while simulating the Naming Game has complex dynamics, and goes through various states before reaching global consensus. We talk about those dynamics as a convergence process, towards the state where all agents share the exact same lexicon, with exactly one word for each meaning without synonymy and homonymy. This state is stable, lexicons will not change anymore whatever are the modalities of the interaction -- which agent is the speaker, which is the hearer, and which meanings and words are used. Convergence and stability for different types of Naming Games has been proved analytically \cite{devylder2007}. In this paper, we do not focus on whether the model converges or not, but on the speed and complexity properties of the dynamics before convergence. Measures for each of those aspects, used to describe the system while in this intermediate state, were defined in previous work \cite{Loreto2011}. We distinguish local measures --accessible to each agent-- from global measures, computed on the whole population.

\subsubsection{TCS: Theoretical Communicative Success}

The Theoretical Communicative Success is a measure of distance to the fully converged state. First, for each meaning, we can consider the probability of having a successful communication when using this meaning as a topic, given a state of the population. The TCS is the average of those probabilities, over all possible meanings. In the case of Random Topic Choice, this measure coincides with the general probability of having a successful interaction. By definition, it is a global measure, not accessible to individual agents. To retrieve its value, we can either estimate it using a snapshot of the population and a Monte Carlo method with random topic choice, or compute it. To detail the exact computation formula, we need to first define the probability of success between two given vocabularies of agents $A$ and $B$. As detailed in the previous section, a vocabulary has 2 components: a coding part, used to find words associated to a meaning, and a decoding part, used to find meanings associated to a word. For vocabulary $V(A)$, we would then have the 2 matrices $V^c(A)$ and $V^d(A)$. If A is the speaker and B the hearer, A is coding and B decoding, hence the formula of the probability of success in this case, averaged over all possible meanings:
\begin{equation}
TCS_s(A,B) = \frac{1}{M}\sum\limits_{m}\sum\limits_{w}V^c(A)_{m,w}\cdot V^d(B)_{m,w}
\end{equation}

Because before an interaction we do not necessarily know which agent will be the speaker and which will be the hearer, the 2 situations (A speaker and B hearer / B speaker and A hearer) are to be considered, as equiprobable. The final value $TCS(A,B)$ is the mean of $TCS_s(A,B)$ and $TCS_s(B,A)$.

To scale up to population level, one can compute an average vocabulary for the whole population $V(P)$, and then the probability of success for an interaction between this lexicon and itself. For a large enough population, this value is indeed a good approximation of the probability of success. $V(P)$ is an element-wise average of the lexicon matrices of all agents.

When using random topic choice, this value abruptly goes from 0 to 1 after a certain number of interactions. These dynamics can be seen on fig.\ref{fig_srtheo}, where the random topic choice is represented -- among active strategies that are explained in a following section.
In practice, we use Monte Carlo estimation for the values at population level over time, and the exact computation for the active topic choice strategy (see following section), as it requires more precision and the population vocabulary is already built.

\subsubsection{Local Complexity\label{local_complexity}}

The starting state of an agent's vocabulary is empty (all-zero matrices), and the end state is identical coding and decoding matrices, with exactly one distinct word per meaning. But between those 2 situations, through which states goes the vocabulary? How much conflictual information (synonymy and homonymy) has to be considered?

For each agent, we can define a local complexity measure, by counting the number of distinct associations present in the vocabulary. In our case, this is exactly the sum of all elements of the matrix $V(A)$. At the beginning of a simulation, while the vocabulary is empty, this measure equals 0. At the end, its value is the number of meanings $M$. When using random topic choice, there is a fast growth to a maximum, before a slow decrease to the final value $M$ (can be seen in fig.\ref{fig_srtheo}). This measure is nearly proportional to the minimal memory needed to represent the lexicon (as a sparse matrix or a list of word-meaning associations), and therefore should remain low in a cognitively plausible situation.

\subsection{Active Topic Choice Strategy}\label{active_strategy}

The main contribution of this work is the definition of an active strategy for the choice of the topic in each interaction, by comparison to the usual choice of picking meanings randomly (with a uniform distribution over the space of meanings). The strategy has to be local, i.e. use only information available to the agent, namely its own vocabulary and results of past interactions it was involved in.

To both converge quickly and control complexity, behavior should be driven by maximization at each interaction of the Theoretical Communicative Success. However, this value is a global measure, therefore not accessible at agent level. Agents only sample information about the global state of the population, or the average vocabulary $V(P)$, through their interactions as hearer or speaker.

The strategies for active topic choice found in previous work are separated in two levels of decision \cite{Schueller2016}. First, a decision between \textit{exploring} a new meaning (that is associated to no words in the vocabulary so far) and choosing (\textit{exploiting}) a meaning among those already used before. Then, if exploiting, deciding which known meaning to use depending on past interaction results.

The strategy introduced in this work keeps those two levels, while basing both decisions on a new measure called Local Approximated Probability of Success (LAPS), using a local representation of $V(P)$.

\subsubsection{LAPS: Local Approximated Probability of Success}

Here, we define an approximation of $V(P)$, $\widetilde{V}(P)$, using information sampled by agents during their interactions. We construct independently the coding and decoding parts $\widetilde{V}^c(P)$ and $\widetilde{V}^d(P)$. For every meaning $m$ (and every word $w$), we use a sliding window over the recent past interactions -- of maximal length $\tau$, the time scale parameter-- and count the number of times it is associated to each word $w'$ (or meaning $m'$). This value divided by $\tau$ is the local estimation of the probability of an other agent coding $m$ using $w'$ (or decoding $w$ as $m'$). With this, we retrieve the values of both matrices $\widetilde{V}^c(P)$ and $\widetilde{V}^d(P)$.

Let $M^c(m)$ be the memory of the past interactions where $m$ was the topic, if there has been $T_m$ such interactions. $w_t$ denotes the word used during the $t^{th}$ interaction of the agent using the meaning $m$. We can now build $\widetilde{V}^c(P)$:
\begin{equation}
M^c(m) = \left( w_{t}\right)_{1\leq t \leq T_m} \hspace{0.6cm} \widetilde{V}^c(P)_{mw} = \frac{\sum\limits_{t=T_m-\tau + 1}^{T_m} \delta_{w,w_t} }{\tau}
\end{equation}

Similarly, by defining $M^d(w)$ be the memory of the past interactions where $w$ was the topic, with $T_w$ such interactions, we can build $\widetilde{V}^d(P)$:
\begin{equation}
M^d(w) = \left( m_{t}\right)_{1\leq t \leq T_w} \hspace{0.6cm} \widetilde{V}^d(P)_{mw} = \frac{\sum\limits_{t=T_w-\tau + 1}^{T_w} \delta_{m,m_t} }{\tau}
\end{equation}

Until $\tau$ interactions have been done with a given meaning or word, $\sum\limits_w\widetilde{V}^c(P)_{mw}$ and $\sum\limits_m\widetilde{V}^d(P)_{mw}$ do not sum to $1$. The remaining probability weight is assumed to be associated with failure.
If we would normalize to $1$, with a single interaction an agent would already estimate as 100\% sure that the same word-meaning association would be used again with the same topic for example. Without the normalization, this happens only after $\tau$ interactions.
In other words, this reflects lack of information due to small sample size.
We define a Local Approximated Probability of Success, or a local equivalent of the Theoretical Communicative Success for an agent $A$ with vocabulary $V(A)$:
\begin{equation}
\label{laps_eq}
LAPS_A = TCS(V(A),\widetilde{V}(P)_{A})
\end{equation}
For some vocabulary update policies called lateral inhibition, similar matrices are computed, but used directly as an agent's own representation of the lexicon. This usage does not prevent the complexity burst \cite{Wellens2012}.

\subsubsection{Exploration vs. Exploitation}
The first choice of our new strategy is between exploring new meanings or exploiting already known ones. Exploration should happen when agents are confident enough about their agreement with the rest of the population over their known meanings \cite{Schueller2016}. The LAPS measure in itself is a measure of confidence, and the simplest way to take this into account is to only explore when reaching the maximum value $\frac{K}{M}$ where $K$ is the number of known meanings and $M$ the total number of meanings in the world. This value can actually be reached, thanks to the sliding window of parameter $\tau$.

\subsubsection{Multi-Armed Bandit}

The second decision process concerns the exploitation part, when picking the topic among the known meanings. We designed a behavior driven by the increase of the LAPS measure. In other words, agents seek the meaning that would yield the greatest increase of LAPS. However, computing the expectancy of this value $\Delta LAPS$ is hard computationally speaking, and therefore not suitable for a model of a cognitive process. We can only consider the process a black box, where following a decision between a finite set of options, a reward value is obtained. This is exactly the definition of the Multi-Armed Bandit problem, associated to a class of reinforcement learning algorithms that have been extensively studied \cite{bubeck2012}.
The name comes from an analogy with a person trying to maximize their gain while facing a set of slot-machines (also called \textit{one armed bandit}), and being able to use only one at a time. The probability distribution of the reward of each machine is unknown, and the player has to both collect information by playing and exploit the highest rewarding machine -- with limited knowledge of its reward distribution -- hence keep balance between exploration and exploitation. In our problem, we can see known meanings as the possible arms, and the reward $\Delta LAPS$. Our case is quite specific, as: 1) distributions are non stationary, 2) they depend on past choices, 3) and the number of arms grows over time (and starts at 0).
This specific situation led us to choose an algorithm, where weights associated to each arm undergo a decay over time, which let them stay at the same order of magnitude of the initial weights of new arms \cite{bclement2015mab}.
Our algorithm depends on 2 parameters: integrated balance between reward-driven exploitation and random exploration between arms through the parameter $\gamma$, and time scale $n$ for the decay of weights. As a reward, we consider the increase of LAPS yielded by the interaction, $\Delta LAPS$, or $0$ if the latter is negative in order to avoid negative weights. See algorithm \ref{algo2}.
\begin{algorithm}
\caption{\textbf{LAPSmax Bandit}\label{algo2} Multi-Armed Bandit algorithm used as a topic choice strategy maximizing the LAPS measure. New arms are created with weights $w_a$ equal to the reward $r_i$ obtained at the end of an interaction with a new meaning.}
\begin{algorithmic}
\REQUIRE $\gamma$ rate of exploration for bandit
\REQUIRE $n$ time scale for weights decay
\REQUIRE vocabularies $V$ and $\widetilde{V}$, \#meanings $M$
\STATE $K \leftarrow \left|V.known\_meanings()\right|$
\IF{$LAPS_A = \frac{K}{M}$}
\STATE $m \leftarrow random\left(V.unknown\_meanings()\right)$
\ELSE
\FOR{$a \in Arms$}
\STATE $\tilde{w}_a = \frac{w_a}{\sum_jw_j}$
\STATE $p_a = (1-\gamma)\cdot\tilde{w}_a + \frac{\gamma}{K}$
\ENDFOR
\STATE Sample $m \in Arms$ using distribution $(p_a)_{a \in Arms}$
\ENDIF
\RETURN $m$
\STATE\{Interact using topic $m$ and compute reward $r$\}
\IF{$m \in Arms$}
\STATE $w_m \leftarrow \frac{n}{n+1}\cdot w_m + r$
\ELSE
\STATE Add $m$ to $Arms$ with $w_m = r$
\ENDIF
\end{algorithmic}
\end{algorithm}
\begin{figure}[t!]
\begin{center}
\includegraphics[width=7.7cm]{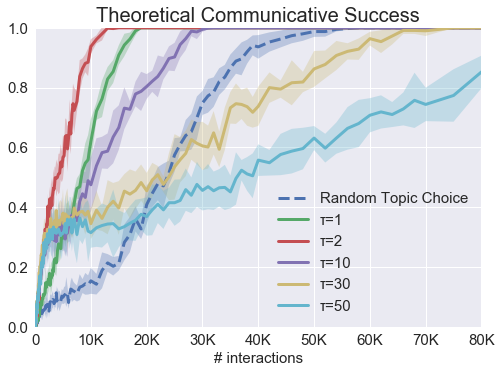}
\includegraphics[width=7.7cm]{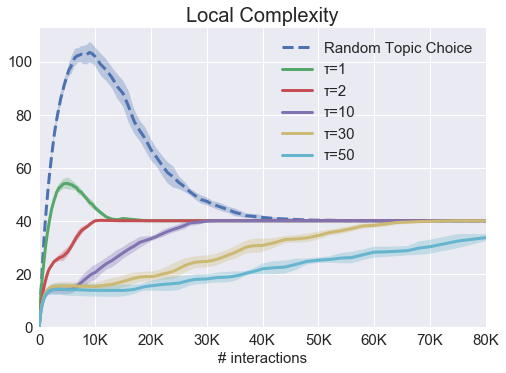}
\end{center}
\caption{Theoretical success rate over time and Local Complexity for Random Topic Choice and Active Topic Choice with several values of the time scale $\tau$ used in the LAPS measure. $N$=$M$=$W$=$40$, $\gamma$=$0.01$, mean over 8 trials.}
\label{fig_srtheo}
\end{figure}
\section{Results}

For all simulations, we set $N$=$M$=$W$=$40$, compute up to 80,000 interactions and take the mean over 8 trials. The situation $M$=$W$ is the most constrained and complex to solve, as synonymy and homonymy are more probable. We ran simulations for $1\leq\tau\leq50$, and set $n$=$\tau$. For the exploration rate, if the condition $0<\gamma\ll1$ is respected, the actual value of $\gamma$ does not matter much, as its only function is to avoid rare cases where some weights reach a value of 0 and cannot be selected anymore. We set $\gamma$=$0.01$. However, we also ran simulations with pure random choices at the bandit level, to be able to study the influence of each level of our algorithm. This case identifies as $\gamma$=$1$.

The evolution of the TCS and complexity over time is represented on fig.\ref{fig_srtheo}, for several values of the time scale $\tau$. They are compared on the same plots with Random Topic Choice. We can see that convergence is faster for low values of $\tau$, the fastest being for $\tau$=$2$, which is 4 times faster than Random Topic Choice. As for complexity, for all configurations excepted $\tau$=$1$ values stay below the final level 40. After reaching a first threshold, they increase linearly with time, the slope being smaller for higher values of $\tau$. For $\tau$=$1$, the maximum value is only half of the maximum reached by Random Topic Choice. It is understandable that $\tau$=$1$ is an outlier: in this case LAPS is an autocorrelation with the current interaction, by definition older interactions are not taken into account.

\begin{figure}[h!]
\begin{center}
\includegraphics[width=7.7cm]{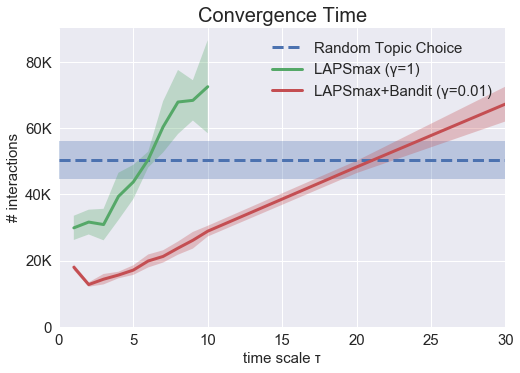}
\end{center}
\caption{Convergence time depending on time scale $\tau$ used in LAPS measure, for both $\gamma$=$0.01$ and $\gamma$=$1$, compared with Random Topic Choice. $N$=$M$=$W$=$40$, mean over 8 trials.}
\label{fig_convtime}
\end{figure}

On fig.\ref{fig_convtime}, we can see the dependency of convergence time on the parameter $\tau$, plotted for configurations $\gamma$=$0.01$, $\gamma$=$1$ and the value for Random Topic Choice as a reference. Both have dynamics consistently faster than Random Topic Choice for low values of $\tau$, however $\gamma$=$0.01$ performs better. Excepted for $\tau$=$1$, convergence time increases linearly with $\tau$ for both, with a minimum at $\tau$=$2$, and a smaller slope for $\gamma$=$0.01$.
\section{Discussion}

Results show that the new strategy presented in this paper 1) allows fast convergence, 2) controls efficiently complexity growth, 3) its dynamics are consistent and highly correlated with parameter change, 4) the 2 levels of the algorithm each contribute to the increased performance.

With $\tau$=2, each agent on average only speaks 15 times about each meaning before convergence (i.e. less than half the population), and information has already been both conveyed between all agents and disambiguated. The linearity of the evolution of TCS and complexity lets think that this algorithm may as well scale efficiently to other values of $N$, $M$ and $W$.
Compared to previous work, this topic choice algorithm is more robust, and optimal parameters are easier to find. It generalized well to Minimal Naming Game and can be used for all other Naming Game models.

LAPS is coherent from a cognitive point of view, and corresponds to an actual internal confidence about quality of communication with the rest of the population.
As stated in the results section, the case $\tau$=$1$ is an outlier, being a simple autocorrelation with the current interaction. The optimal value $\tau$=$2$ is then the lowest possible value taking into account past interactions, i.e. $\widetilde{V}(P)$ takes the lowest possible memory, which is therefore credible for humans. Further work will be needed to determine for which values of N, M and W $\tau$=2 stays the optimal value.

\section*{Acknowledgments}

The IdEx program (Univ. de Bordeaux) allowed W. Schueller to visit V. Loreto. We thank Miguel Iba\~{n}ez Berganza and Benjamin Cl\'{e}ment for the fruitful discussions.

\subsubsection*{Source code}

The Python code used for the simulations of this paper is available as open source software: \textbf{\href{https://github.com/wschuell/notebooks\_cogsci2018}{https://github.com/wschuell/notebooks\_cogsci2018}}

\bibliographystyle{apacite}
\setlength{\bibleftmargin}{.125in}
\setlength{\bibindent}{-\bibleftmargin}

\bibliography{biblio}

\end{document}